%% file: main_PRL_version.tex
\documentclass[aps, prl, reprint, superscriptaddress, nobibnotes, amsmath, amssymb]{revtex4-2}

\usepackage{graphicx}
\usepackage{url}
\usepackage{textcase}
\usepackage{bm}
\usepackage{xcolor}
\usepackage{arydshln}
\usepackage{braket}
\usepackage{svg}
\usepackage{ulem}
\usepackage{hyperref}
\hypersetup{
  colorlinks = true,
  linkcolor = blue,
  urlcolor = blue,
  citecolor = blue
}
\usepackage{comment}
\usepackage{soul}
\usepackage{amsmath}

\begin{document}

\include{definitions}

\title{Coherent Ultrafast Excitonic Oscillations in Monolayer WS$_2$}

\author{Jorge Cervantes-Villanueva}
\affiliation{Institute of Materials Science (ICMUV), University of Valencia,  Catedr\'{a}tico Beltr\'{a}n 2,  E-46980,  Valencia,  Spain}

\author{Alberto Garc\'ia-Crist\'obal}
\affiliation{Institute of Materials Science (ICMUV), University of Valencia,  Catedr\'{a}tico Beltr\'{a}n 2,  E-46980,  Valencia,  Spain}

\author{Davide Sangalli}
\affiliation{Istituto di Struttura della Materia-CNR (ISM-CNR)  and European Theoretical Spectroscopy Facility (ETSF), Piazza Leonardo da Vinci 32, 20133 Milano, Italy}

\author{Alejandro Molina-S\'{a}nchez}
\affiliation{Institute of Materials Science (ICMUV), University of Valencia,  Catedr\'{a}tico Beltr\'{a}n 2,  E-46980,  Valencia,  Spain}

\date{\today}

\begin{abstract}

Monolayer transition metal dichalcogenides are a suitable platform for studying excitonic coherence in the light–matter coupling regime. We present an ab initio time-dependent GW–Bethe–Salpeter equation (GW–BSE) investigation of coherent excitonic dynamics in monolayer WS$_2$. By solving the coherent coupling between the A, A$^{*}$, and B excitons under linearly polarized pump fields, we identify the microscopic origin of the resulting oscillatory dynamics and rationalize it using an effective theoretical model. Our results provide the interpretation of recently reported coherent excitonic phenomena in monolayer WS$_2$ (Nano Lett. 24, 8117 (2024)). Building on this first-principles time-resolved framework, we propose a tailored pump–probe scheme that enables the controlled generation and regeneration of coherent oscillations between excitonic states. These findings establish a predictive route for controlling excitonic coherence in two-dimensional materials, with direct relevance for ultrafast optoelectronic switches and solid-state quantum logic devices.

\end{abstract}

\maketitle

\textit{Introduction}---Optical coherence phenomena are fundamental in understanding light-matter interactions and have potential applications in quantum technologies, using coherent control of qubits \cite{warner2025,Cai2023}. Among these phenomena, excitonic Rabi oscillations are responsible for the coherent light-matter interaction in semiconductors \cite{Rabi2001,Nguyen2023}. They are especially outstanding in two-dimensional (2D) materials, such as monolayer transition metal dichalcogenides (TMDs), which exhibit strong excitonic effects \cite{Mak2016,Selig2016,PhysRevLett.105.136805,Mueller2018,Ugeda2014}. They manifest large exciton binding energies and strong light–matter coupling due to small 2D dielectric screening and quantum confinement effects \cite{RevModPhys.90.021001}.

Exciton dynamics and their coherence phenomena are typically investigated using ultrafast pump–probe spectroscopy \cite{Hao2016,Trovatello2020,Bae2022,Moody2015,Trovatello2022}. These studies have revealed that exciton formation in TMDs occurs on the order of hundreds of femtoseconds \cite{Karni2022}, while coherent spin-valley exciton dynamics can persist up to nanoseconds \cite{Xu2014}. Such long-lived coherence makes these materials attractive for quantum information processing \cite{Liu2019}. The coupling between spin, valley, and excitonic degrees of freedom in monolayer TMDs offers a versatile platform for exploring novel physical phenomena and developing future optoelectronic and quantum devices \cite{brotons-gisbert_spinlayer_2020,Koperski2015,Schaibley2016}.

In this context, recent pump–probe reflectance experiments have demonstrated the presence of excitonic Rabi oscillations in monolayer tungsten disulphide (WS$_{2}$) under non-resonant excitation conditions  \cite{Timmer2024}. These experiments suggest that simultaneous population of A and B excitonic states can generate rapid coherent oscillations with a period of 11.5 fs, arising from their mutual coupling during the exciton coherence time (on the order of tens of femtoseconds). However, the complex nature of excitons in 2D materials, which involves a rich spectrum of bound states, necessitates the use of predictive ab initio approaches, especially when modeling ultrafast dynamics \cite{PhysRevLett.128.016801,Chan2023,PhysRevB.84.245110,gosetti_unveiling_2025,Chan2025,Calandra2025}. 

In this Letter, we present a first-principles approach for simulating pump–probe experiments, including pump and probe fields on equal footing. Our approach accurately reproduces experimental conditions and allows for a detailed analysis of exciton dynamics and coherent excitonic phenomena. Our simulations confirm the existence of coherent oscillations between the A and B excitons, but reveal more complex coherent dynamics than previously captured by models involving only a few excitonic states \cite{Timmer2024}. We identify the critical role of the A$^{*}$ state, which lies energetically between the A and B excitons, in driving multi-level coherent dynamics. We elucidate the origin of this coupling using an effective three-level model. Based on this analysis, we propose a tailored pump–probe scheme that enables the on-demand generation and regeneration of coherent oscillations. Our findings not only characterize the fundamental coherent coupling between the main resonances in monolayer WS$_2$, but also establish an all-optical protocol with direct applications for ultrafast optoelectronic switches and solid-state quantum logic \cite{timmer_ultrafast_2026}.

\textit{Theoretical approach}---Theoretical pump–probe simulations are carried out within the framework of many-body perturbation theory (MBPT). The equilibrium electronic and optical properties are computed within the DFT+GW+BSE scheme. The time evolution of the system is modeled based on the equation of motion (EOM) for the time-dependent one-body density-matrix projected into the single particle basis ~\cite{PhysRevMaterials.5.083803}:

\begin{equation}\label{Eq(1)}
\partial_{t} \rho_{nm\mathbf{k}}(t) = - i[ h^{\textrm{HSEX}}(t), \rho(t)]_{nm\mathbf{k}}  -\eta \rho_{nm\mathbf{k}}(t) \\
\end{equation}
with
\begin{equation}
 h^{\textrm{HSEX}} = h^{\textrm{QP,eq}} + \Delta \Sigma^{\text{HSEX}} + U^{\textrm{ext}}(t)
\end{equation}

Here and in the following, we use atomic units. The term $h^{\textrm{QP,eq}}_{nm\mathbf{k}}=\delta_{n,m}\epsilon_{n\mathbf{k}}$ contains the equilibrium quasiparticle energies obtained from GW calculations, while $\Delta \Sigma^{\text{HSEX}}$ represents the variation of the many-body self-energy within the Hartree plus screened exchange (HSEX) approximation, where the screening is constant and takes the value at equilibrium. The term $U^{\text{ext}}(t)$ accounts for the interaction with external fields (pump and probe pulses) such that the entire time evolution of the system arises from their influence. Relaxation and dissipation mechanisms can be incorporated in the equation of motion via the collision integral, $I(t)$ \cite{Sangalli2018,PhysRevLett.128.016801}. Because our focus is strictly on the ultrafast coherent dynamics immediately following excitation (a regime where incoherent electron-phonon and carrier-carrier scattering have not yet fully thermalized the system \cite{Caruso2021}) we approximate the collision integral with a phenomenological constant dephasing rate, $\eta$. This term effectively captures the decoherence of the system on these sub-picosecond timescales. The transient absorption signal, $\chi(\omega)$, is obtained by applying a Fourier transform to the time-dependent polarization $P(t) = P_{P_{p}}(t) - P_{P}(t)$, where $P_{P_{p}}(t)$ and $P_{P}(t)$ represent the polarizations calculated from simulations including both pump and probe pulses, and the pump pulse alone, respectively \cite{PhysRevB.89.064304,PhysRevB.92.205304}. Both the pump and probe pulses are modeled using quasi-sinusoidal envelopes, which allow precise control of the relative phase between them (see the SI for details on the MBPT and EOM calculations).

\begin{figure}
\includegraphics[width=1.0\linewidth]{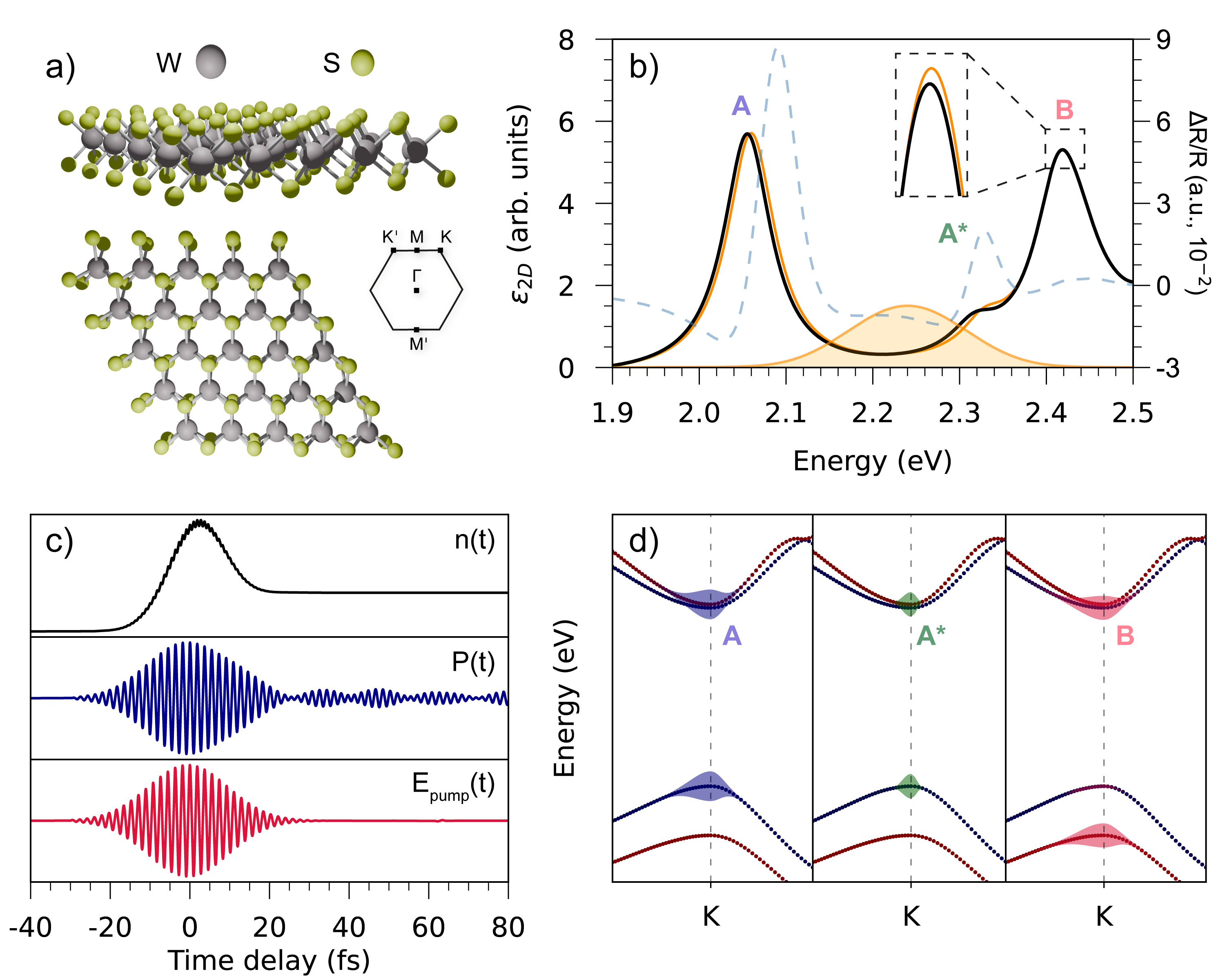}
\caption{\label{Estructure} 
Pump-probe setup and excitonic response in monolayer WS$_{2}$ with pump at 2.24 eV and pump-probe time delay of 50 fs. (a) Side and top views of the monolayer WS$_{2}$ crystal structure, along with its Brillouin zone. (b) Absorption spectra at equilibrium (black line), out-of-equilibrium after the pump (orange line) and the resulting $\Delta$R/R signal (blue dashed line). The pump pulse (orange shaded area) is detuned equally from the A and B excitonic resonances. (c) Time evolution of the excited density (black), polarization (blue), and pump pulse (red). (d) Electronic band structure of WS$_{2}$ showing the A, A$^{*}$, and B resonances, and the corresponding bands that contribute to their formation.}
\end{figure}

\textit{Results}---We investigate monolayer WS$_2$ (see crystal structure in Fig~\hyperref[Estructure]{1(a)}), along with its equilibrium absorption spectrum (black line in Fig.~\hyperref[Estructure]{1(b)}). The spectrum exhibits three main excitonic resonances: A at 2.06 eV, A$^{*}$ at 2.32 eV, and B at 2.41 eV, in agreement with prior theoretical and experimental studies \cite{PhysRevB.103.155152,Timmer2024}. The presence of the A$^{*}$ resonance introduces richer excitonic dynamics, resembling conditions at low temperatures. However, in recent works focused on coherent oscillations between the A and B resonances, the A$^{*}$ state is excluded since its signal is suppressed at room temperature \cite{Zhu2015}. To excite both A and B resonances, the pump field is centered at 2.24 eV, as shown by the orange shaded region in Fig.~\hyperref[Estructure]{1(b)}. The pump has a duration of 10 fs and a peak intensity of $5\times10^{5}$ kW/cm$^2$, ensuring strong excitation ($\sim 5\times10^{10}$ cm$^{-2}$) despite the energy detuning. The probe pulse is broader (1 fs) and weaker ($1\times10^{2}$ kW/cm$^2$), enabling it to probe the entire spectrum without significantly altering the system. This configuration yields the out-of-equilibrium absorption spectrum and the corresponding $\Delta$R/R signal (orange and dashed blue lines in Fig.~\hyperref[Estructure]{1(b)}, respectively) for a representative 50 fs pump–probe delay. The slight blue shift in the resonance energies observed in the $\Delta$R/R spectrum arises from the use of equilibrium dielectric screening in the calculations \cite{PhysRevMaterials.5.083803}. Fig.~\hyperref[Estructure]{1(c)} presents the time evolution of the excited density (black), macroscopic polarization (blue), and the pump field (red). The persistence of both a residual excited population and an oscillating macroscopic polarization long after the pump pulse has passed is a direct signature of the successful preparation of coherent excitonic states \cite{Gosetti2024}. Finally, Fig.~\hyperref[Estructure]{1(d)} shows the band structure origin of the A, A$^{*}$, and B resonances; an aspect that will be central to the discussion of the coherent oscillations in the following section. The A and A$^{*}$ excitons arise from the same electronic bands, with A$^{*}$ exhibiting a much smaller oscillator strength \cite{RydbergSeries_WS2}.

\begin{table}[!b]
\caption{\label{Energy_splitting} Oscillation frequency components and energy splittings for different pairs of resonances.}
\begin{ruledtabular}
\begin{tabular}{lccc}
{Coupling} & {Osc. (fs)} & {Osc. (meV)} & {E. spliting (meV)} \\ \hline
 {A - A$^{*}$} & 15.5 & 267   & 270  \\
 {A - B} & 8.14 & 508   & 363 \\
 {A$^{*}$ - B} & --- & ---  & 90 \\
\end{tabular}
\end{ruledtabular}
\end{table}

\begin{figure}[t!]
\includegraphics[width=1.0\linewidth]{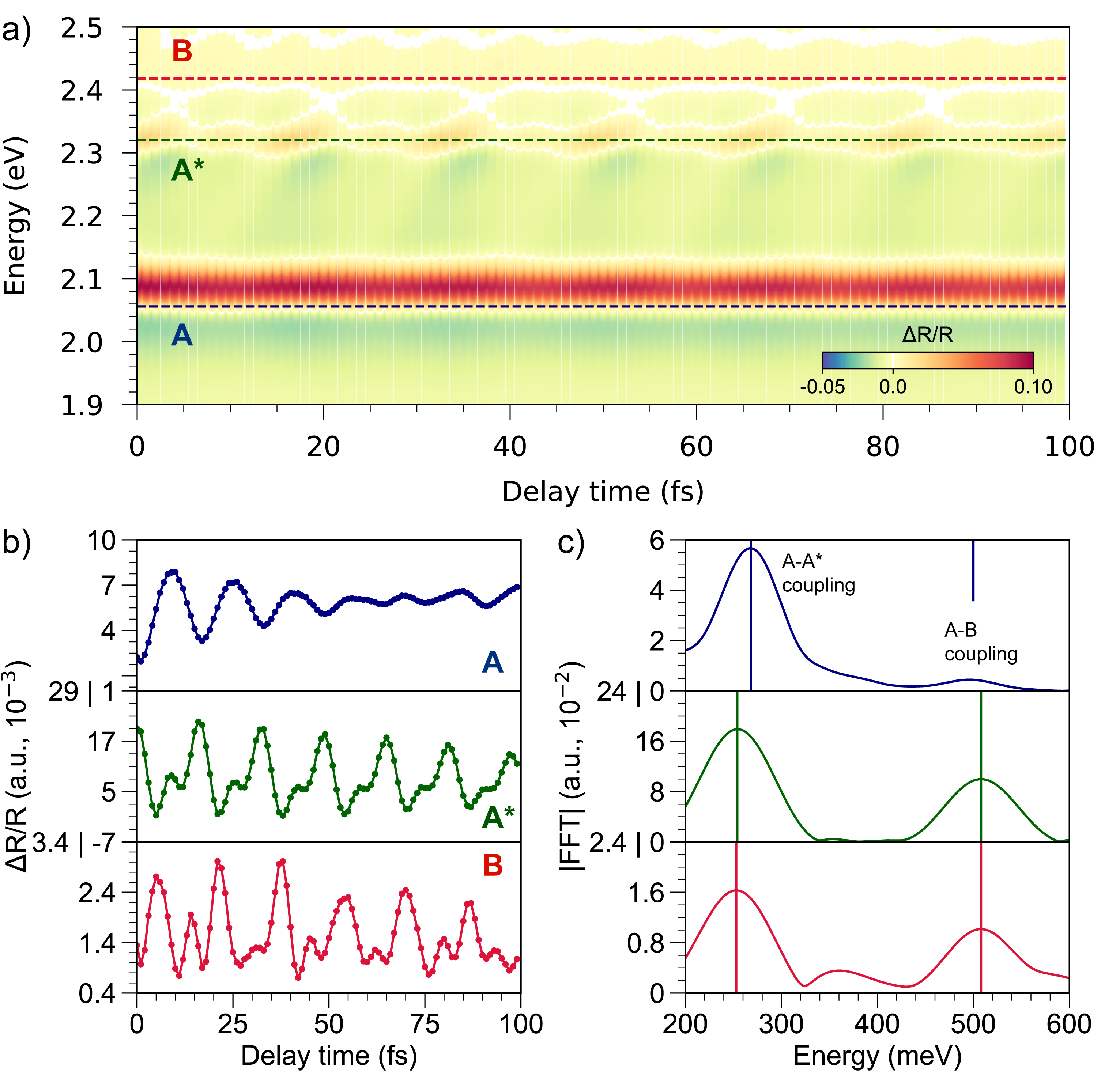}
\caption{\label{Colormap} Ultrafast pump-probe simulations on monolayer WS$_{2}$. (a) Pump-probe map of $\Delta$R/R signal. Resonances A, A$^{*}$ and B are highlighted in blue, green and red, respectively. (b) Time-domain coherent oscillations and (c) their corresponding frequency components for the A, A$^{*}$ and B resonances.}
\end{figure}

We apply our method to investigate the coherent exciton dynamics in the system. Pump-probe calculations as a function of the time delay are depicted in Fig.~\hyperref[Colormap]{2(a)}, where the dynamics of the resonances defined in Fig.~\hyperref[Colormap]{1(b)} are captured. Coherent oscillations are observed in both A and B excitonic resonances, due to their mutual coupling and consistent with previous reports in the literature \cite{Timmer2024}. However, the emergence of the additional A$^{*}$ resonance introduces significant modifications to the coherent exciton dynamics, leading to a richer and more complex case. In Fig.~\hyperref[Colormap]{2(b-c)} are displayed the time-dependent oscillations of the three resonances, along with their corresponding frequency components. Table \ref{Energy_splitting} summarizes the characteristic oscillation frequencies and energy splittings associated with the different pairs of resonances. Coherent dynamics are primarily governed by two dominant frequency components: a lower energy mode at approximately 265 meV, and a higher energy mode near 510 meV. These frequencies correlate with the energy splittings between excitonic resonances, being about 270 meV for the A-A$^{*}$ splitting and close to 363 meV for the A-B case. 

To investigate the origin of the coherent oscillations, we have developed a three-level model based on the evolution of the density matrix governed by the master equation in Lindblad form $\frac{d \Gamma(t)}{d t} = - i [H(t), \Gamma(t)]$, where $\Gamma(t)$ is the many-body density matrix. $H(t)$ represents the many-body Hamiltonian of the system, while relaxation and decoherence effects are neglected by setting Lindblad terms to zero, i.e., $\mathcal{L}(\Gamma(t)) = 0$. If we define the ground state $\ket{0}$, and two excited states $\ket{A}$ and $\ket{B}$, representing the excitonic resonances, we obtain:

\begin{align}\label{Eq_model}
\Gamma_{AB}(t) \sim (\Gamma_{BB}(t) - \Gamma_{AA}(t)) \frac{\Omega_{AB}(t)}{2} + \notag \\ 
+\frac{\Omega_{OA}(t)}{2} \Gamma_{0B}(t) - \frac{\Omega_{0B}(t)}{2} \Gamma_{A0}(t)
\end{align}

where $\Omega_{ij}(t)$ are the Rabi frequencies. $\Omega_{AB}(t)$ describes the coupling between A and B excitons, and it can be formulated in term of exciton-exciton dipoles~\cite{Sangalli2023}. The model reveals that coherent oscillations between two excited states originate from an imbalance in their excitonic populations. When excitonic populations associated with both resonances coexist coherently but exhibit an amplitude imbalance, a coherent coupling is manifested, giving rise to observable oscillations in the system dynamics (see SI for a detailed explanation of the model). This interpretation can be extended to the multi-resonances coherent dynamics presented in Fig.~\hyperref[Colormap]{2}. The excitonic populations generated in the different states are not identical due to their distinct transition strengths and dephasing dynamics. Such population asymmetry satisfies the condition identified by the model, thereby enabling the emergence of coherent oscillations between the excitonic states. 

\begin{figure*}[!t]
\label{Generator}
\includegraphics[width=1.0\linewidth]{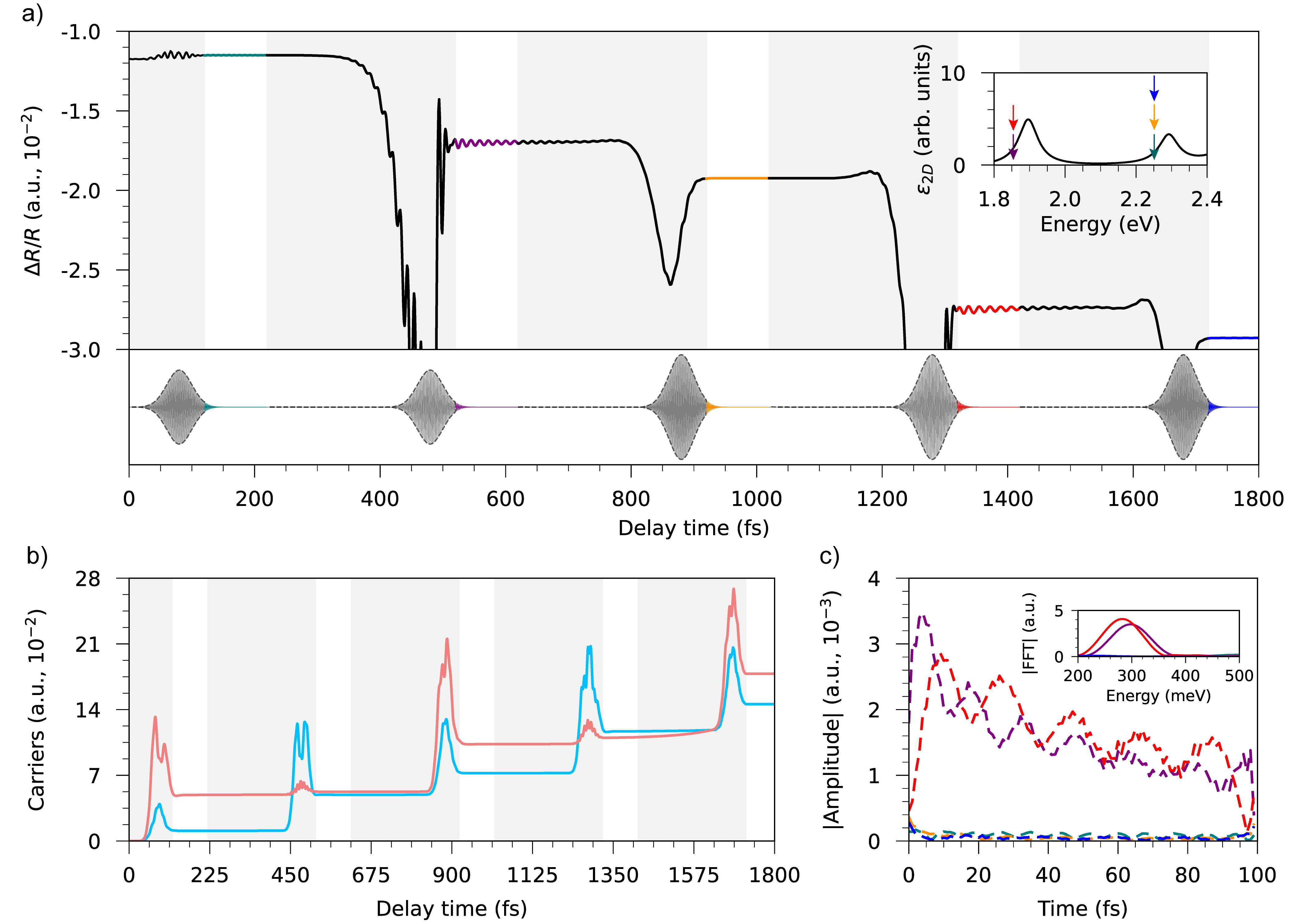}
\caption{Pump-probe configuration for realizing a generator of coherent oscillations in monolayer WS$_{2}$. (a) $\Delta$R/R signal monitored at the A resonance, shown alongside the temporal profiles of the applied pump pulses. Shaded regions indicate the preparation regime before the signal stabilizes, while the colored lines highlight the time windows relevant for the generator operation. The inset shows the pump pulse energies to excite resonantly resonances A and B. (b) Time evolution of the excited population densities of the A (blue) and B (red) resonances. The coherence term $\Gamma_{AB}$ develops when pump A (pump B) acts on the system in presence of a coherent B state (A state) previously induced by the action of pump B (pump A). (c) Amplitudes of the $\Delta$R/R signal in the different generator configurations. The coherent oscillations exhibit an amplitude approximately one order of magnitude larger than in the non-oscillatory regime. The inset depicts the frequency components of the signals, illustrating that the dominant oscillation frequency matches the energy splitting between the A and B excitonic resonances.}
\end{figure*}

Fig.~\hyperref[Colormap]{2(c)} shows oscillations dominated exclusively by the low-frequency mode at the A exciton energy. This behavior is attributed to the higher excitation densities associated with the A and A$^{*}$ resonances compared to B and to the relatively large spectral separation between A and B. In contrast, oscillations at A$^{*}$ exhibit two prominent frequency components. The lower energy component corresponds to the same A-A$^{*}$ transition energy. The appearance of the higher frequency mode, linked to the A-B coupling, stems from the fact that both A and A$^{*}$ involve transitions between the same electronic bands, as shown in Fig.~\hyperref[Estructure]{1(d)}. Their spectral proximity further enhances their interaction.  It is important to highlight that the frequency associated with the A-B oscillations differs from the actual energy splitting between these resonances. This discrepancy arises due to the presence of the A$^{*}$ resonance, which alters the effective coupling dynamics.

This interpretation is supported by additional pump-probe simulations performed using two different \textbf{k}-point grids: a coarse $12 \times 12$ \textbf{k}-grid, where the A$^{*}$ resonance is absent and a finer $36 \times 36$ \textbf{k}-grid, where it appears. The comparison (see SI) confirms that the inclusion of the A$^{*}$ resonance significantly affects the coherent response of the B state. The coherent oscillations of the B resonance follow a pattern similar to those of A$^{*}$, further reinforcing the impact of the A$^{*}$ state on the dynamics of the system. In the recent experimental study \cite{Timmer2024}, the measurements were performed at room temperature, where the A$^{*}$ exciton was thermally suppressed. As a result, the observed oscillation frequency matched the expected two-level coupling scenario. In contrast, our simulations correspond to the zero-temperature limit, allowing the A$^{*}$ state to remain active. The absence of thermal suppression leads to multi-level coupling dynamics, which breaks the frequency matching observed experimentally.

In the pump-probe experiment shown in Fig.~\hyperref[Colormap]{2}, the coherent dynamics are driven by a single pump pulse that is equally resonant with A and B resonances. However, since the coherent oscillations originate from the coherent population transfer from the A to the B exciton (and viceversa), the coherent oscillations can, in principle, be actively controlled through an appropriately designed pump-probe configuration. By tailoring the spectral width, temporal delay, and relative intensities of the pulses, one can selectively engineer the population imbalance and thereby tune the amplitude and visibility of the resulting coherent oscillations. In this way, a pump-probe setup can be configured to operate as a controllable source of coherent oscillations. This analysis is performed on a coarser $12 \times 12$ $\mathbf{k}$-grid (rather than the converged $72 \times 72$ $\mathbf{k}$-grid used in earlier simulations) in order to isolate the A and B resonances while effectively suppressing the higher energy A$^{*}$ state, consistent with experimental conditions at room temperature \cite{Timmer2024}. Although the spectral positions are not fully converged in this setup, this does not impact the qualitative outcome; the ability to generate coherent oscillations arising from the coherent coupling between the two excitonic states. While this demonstrates the viability of the generator under standard room-temperature operating conditions, our earlier findings imply that operating such a device at cryogenic temperatures would activate the $A^*$ state, opening pathways for more complex, multi-level coherent control.

The generation mechanism of the coherent oscillations is illustrated in Fig.~\hyperref[Generator]{3(a)}, which displays the $\Delta$R/R signal monitored at the A resonance for a tailored sequence of linearly polarized pump pulses. The corresponding central energies are shown in the inset of Fig.~\hyperref[Generator]{3(a)}. Each pump has a duration of 20 fs. The first two pulses operate at an intensity of 5×10$^6$ kW/cm$^2$, while the subsequent three are increased to 1×10$^7$ kW/cm$^2$ to enhance the amplitude of the regenerated coherent oscillations. These parameters are explicitly chosen to selectively address the targeted transitions and drive a substantial excitonic population. 

To actively control the coherent coupling, we manipulate the individual excitonic populations. Initially, a pump pulse centered at 2.24 eV selectively excites the B resonance, exciting, from the ground state, a coherent excitonic B state, however without driving an oscillatory response in the probe signal, as shown in Fig.~\hyperref[Generator]{3(b)}. Subsequently, a 1.84 eV pump excites the A resonance, exciting, from the ground state, a coherent excitonic A state that is eventually comparable in magnitude with the previous one. Moreover, due to the $\Omega_{AB}$ term in Eq.~\ref{Eq_model} the second pump also induces a direct excitation from the A to the B exciton \cite{Cha2016}, triggering the onset of coherent oscillations. To demonstrate the viability of this platform as an on-demand coherent-oscillation generator, we apply subsequent pulse sequences to repeatedly initialize the quantum state. By applying a third pump at 2.24 eV, we replenish the population of the B resonance, thereby re-establishing the necessary population imbalance. A subsequent 1.84 eV pulse addressing the A resonance immediately restores the coherence, re-initiating the coherent oscillations with dynamics identical to the first cycle, as depicted in Fig.~\hyperref[Generator]{3(b)}. A final 2.24 eV pump confirms that this deterministic regeneration can be reliably repeated. The functionality of this coherent control persists robustly as long as the system remains within the linear optical regime. To further confirm the distinction between oscillatory and non-oscillatory regions, the corresponding oscillation amplitudes and frequency components are presented in Fig.~\hyperref[Generator]{3(c)} and its inset, respectively. The coherently driven regions exhibit amplitudes more than an order of magnitude larger than those of the non-oscillatory intervals.  

Hence, we demonstrate a coherent-oscillation generator based on a tailored pump-probe configuration in monolayer WS$_2$. In principle, this mechanism can be extended to other TMDs using appropriately designed excitation schemes. We emphasize that the implementation of this approach requires high-quality samples to ensure sufficient coherence control and oscillation resolution. Such sample quality has already been achieved in monolayer TMDs encapsulated in hexagonal boron nitride (hBN) \cite{PhysRevX.7.021026}. The controlled generation and regeneration of coherent exciton oscillations constitute a key step toward coherence-based quantum technologies, opening pathways to ultrafast all-optical switches and solid-state quantum logic devices \cite{sortino_light-matter_2025,pareek_driving_2026}.

\textit{Conclusions}---We have investigated the coherent excitonic dynamics of monolayer WS$_{2}$ by calculating the equilibrium and out-of-equilibrium optical properties within the MBPT framework, applying the EOM of the time-dependent density matrix. Our ab initio simulations reveal that coherent oscillations arising from the coupling between the A and B excitonic resonances are present, consistent with recent experimental observations. Importantly, we find that dynamics are significantly influenced by the presence of an additional resonance, A$^{*}$, which modifies the nature of the coherent coupling. Based on these findings, we propose a pump-probe configuration enabling the on-demand generation and regeneration of coherent oscillations, applicable in any suitable material as TMDs or halide perovskites \cite{qin_coherent_2025}. Moreover, the excitonic oscillations falls in a time scale observable with TR-ARPES under suitable experimental conditions \cite{madeo_directly_2020}. This work provides a comprehensive theoretical framework for understanding coherent exciton coupling in 2D materials and establishes monolayer WS$_{2}$ as a promising platform for generating controlled coherent oscillations, with potential applications in quantum coherence-based devices \cite{Turunen2022,coherent_device_2003}.

\textit{Acknowledgments}---This work is supported by the Horizon Europe research and innovation program of the European Union under the Marie Sklodowska-Curie grant agreement 101118915 (TIMES). This work is part of the project I+D+i PID2023-146181OB-I00 UTOPIA, funded by MCIN/AEI/10.13039/501100011033, the project PROMETEO/2024/4 (EXODOS) and SEJIGENT/2021/034 (2D-MAGNONICS) funded by the Generalitat Valenciana. This study is also part of the Advanced Materials program (project SPINO2D), supported by MCIN with funding from the European Union NextGenerationEU (PRTR-C17.I1) and Generalitat Valenciana. Authors thankfully acknowledges the computer resources at Agustina and technical support provided by BIFI and Barcelona Supercomputing Center (FI-2025-2-0001), and the computer resources at Tirant-UV (project lv48 - FI-2025-2-0001). DS aknowledges funding from the ``MAterials design at the eXascale'' (MaX) centre of excellence co-funded by the European High Performance Computing joint Undertaking (JU) and participating countries (Grant Agreement No. 101093374), and from PRIN project ``Exploring extreme ultraviolet excitons with attosecond time resolution'' (EXATTO), Grant No. 2022PX279E from MIUR (Italy).

\textit{Data availability}---The data that support the findings of
this article are openly available \cite{cervantes_villanueva_2026_19251393}.

\bibliographystyle{apsrev4-2}

%

\end{document}

%% file: definitions.tex
\newcommand{\revtex}{REV\TeX\ }
\newcommand{\classoption}[1]{\texttt{#1}}
\newcommand{\macro}[1]{\texttt{\textbackslash#1}}
\newcommand{\m}[1]{\macro{#1}}
\newcommand{\env}[1]{\texttt{#1}}
\ExplSyntaxOn
\NewDocumentCommand{\longdash}{ O{2} }
 {
  --\prg_replicate:nn { #1 - 1 } { \negthinspace -- }
 }
\ExplSyntaxOff
\setlength{\textheight}{9.5in}

\newcommand{\editor}[2]{%
  \expandafter\newcommand\csname #1note\endcsname[1]{%
    \textcolor{#2}{(\textbf{#1:} \it ##1)}}%
  \expandafter\newcommand\csname #1\endcsname[1]{%
    \textcolor{#2}{##1}}%
  \expandafter\newcommand\csname #1cancel\endcsname[1]{%
    \textcolor{#2}{\sout{##1}}}%
  \expandafter\newcommand\csname #1change\endcsname[2]{%
    \textcolor{#2}{\sout{##1} ##2}}%
  \newenvironment{#1text}{\color{#2}}{\color{black}}
}

\editor{DS}{blue}

\def \ai{\textit{ab initio}~}
\def \WSe{WSe$_{2}$~}
\def \wse{WSe$_{2}$}
\setcounter{secnumdepth}{3}
\renewcommand{\thesubsection}{\Alph{subsection}}